\documentclass[dvips]{article}

\usepackage{icrc2011}

\title{The Upgrade of VERITAS with High Efficiency Photomultipliers}

\shorttitle{N.\ Otte for the VERITAS Collaboration, VERITAS upgrade with high efficiency PMTs}

\authors{A. Nepomuk Otte$^{1}$ for the VERITAS Collaboration$^{2}$,\\ and  L.\
Gebremedhin$^{1}$, K.\ Kaplan$^{1}$, and D.\ Long$^{1}$}
\afiliations{$^1$SCIPP, University of California, 1156 High Street, Santa Cruz , CA 95060, U.S.A.\  \\
$^2$see J. Holder et al. (these proceedings) or http://veritas.sao.arizona.edu/conferences/authors?icrc2011}
\email{nepomuk.otte@gmail.com}

\abstract{We are in the process of upgrading the VERITAS array of Cherenkov telescopes with new, high efficiency photomultipliers (PMT) that will considerably lower the energy threshold of the instrument and improve the overall sensitivity. The upgrade will be finished in Summer 2012 when the PMTs will be installed in the existing cameras. We discuss the performance of the new photon detectors and the status of the project. }

\begin{document}
\maketitle

\section{Introduction}

By means of gamma-ray observations with the latest generation of imaging atmosphere Cherenkov telescopes (IACTs), much insight has been gained into the non-thermal processes in a variety of astrophysical sources, for example, active galactic nuclei, supernova remnants, and  pulsars. Observations in gamma-rays above 100\,GeV also provide a tool to study interesting questions that are of fundamental importance in physics and cosmology. Some examples are the nature of dark matter and the density and evolution of the extragalactic background light.

VERITAS, the Very Energetic Radiation Imaging Telescope Array System, is an array of four imaging atmospheric Cherenkov telescopes located in southern Arizona, USA. Each of the four telescopes has a Davies-Cotton arrangement of 350 identical, hexagonal mirror facets yielding a 12\,m diameter collector with  f/D=1. Located in the focal plane of each telescope is a pixelated camera consisting of 499 photomultiplier tubes (PMTs), each with an angular size of 0.15 degrees \cite{jamie}.

The VERITAS array has achieved, and in some metrics surpassed, its design specifications and science goals; however, the performance parameters are not limited by any fundamental physical constraints. Improved performance can be achieved through improved data analysis techniques and hardware upgrades, which will allow us to better address the key science goals. The ongoing upgrade program of VERITAS, includes:

 \begin{enumerate}
 \item relocating one of the telescopes,
 \item implementing a different hardware trigger scheme, and
 \item refitting the telescope cameras with more sensitive photo-detectors.
 \end{enumerate}

 \begin{figure}[!t]
  \vspace{5mm}
  \centering
  \includegraphics[width=3.2in]{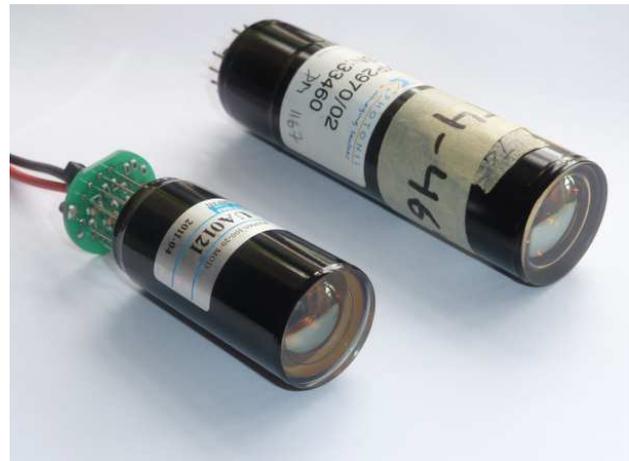}
  \caption{In front the R10560-100-20 MOD from Hamamatsu, which is the replacement for the  XP2970 from Photonis, shown in the back.}
  \label{pmtfig}
 \end{figure}

The relocation of one of the telescopes was completed in summer 2009, which improved the sensitivity by about 30\% \cite{jeremy} and lowered the energy threshold by about 10\%. The upgrade with a new trigger system is ongoing and will be completed this summer. Details of the trigger upgrade can be found in \cite{amanda}. In this paper we discuss the replacement of the photon detectors in the VERITAS cameras with high efficiency PMT, with the goal to lower the threshold to below 100\,GeV and improve the overall sensitivity of VERITAS. 

\section{Overview of the Upgrade Photomultiplier R10560-100-20 MOD}

The selection and evaluation process of photon-detector candidates for the upgrade of the VERITAS cameras began in 2008 and was finished in 2010 with the selection of the photomultiplier R10560-100-20 MOD from Hamamatsu, hereafter R10560. Figure \ref{pmtfig} shows the R10560 in comparison to the Photonis XP2970 that is currently used in VERITAS. The R10560 is a one inch photomultiplier tube with a UV glass entrance window. It is one of the relatively new PMTs with a superbialkali photocathode that features a considerable higher quantum efficiency than classical bialkali photocathodes.

The R10560 has a linear focused dynode structure with eight stages that is biased in a 4, 1, 1, 1, 1, 1, 1, 1, 1 configuration. The voltage divider consists of a passive divider chain between cathode and dynode 5. Between dynode 6 and the anode the bias is stabilized by an active divider circuit. In this configuration the nominal gain is $2\cdot10^5$  at a bias voltage of 1100\,V. In VERITAS the R10560 will be operated at a gain of  $2\cdot10^5$, the same gain as the XP2970 is operated now.

\subsection{Single Photoelectron Signal Characteristics}

 \begin{figure}[!ht]
  \vspace{5mm}
  \centering
  \includegraphics[width=3.in]{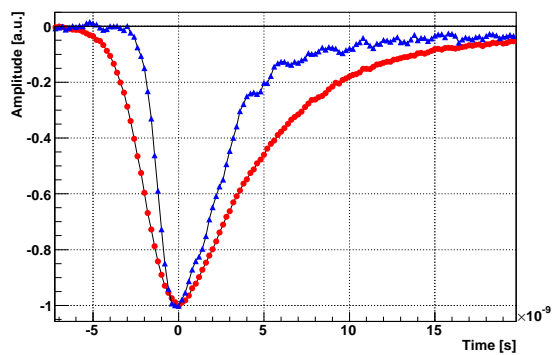}
  \caption{Blue triangles: pulse shape of the R10560. Red solid dots: Pulse shape of the XP2970. Each curve is the average of a number of afterpulses. The individual afterpulses were normalized to the same amplitude before averaging.}
  \label{pulseshape}
 \end{figure}

Figure \ref{pulseshape} compares, normalized to equal amplitudes, afterpulses of an XP2970 and an R10560 at the input of the digitizer boards in VERITAS. The measurement was made on PMTs installed in VERITAS. I.e.\, included in the signal shapes is the dispersion due to 185 ft of RG59 coaxial cable between the camera and the readout at the base of the telescope. Afterpulses can be considered as a perfect pile up of photoelectrons and thus their signal is that of a single photoelectron signal. The single photoelectron signal of the R10560 has a full width at half maximum (FWHM) of 4.2\,nanoseconds, which is about 40\% narrower than the pulse shape of the Photonis XP2970 (6.8\,nanoseconds).   The narrower pulse shape of the R10560 will help to better discriminate the Cherenkov light from air showers against fluctuations in the night sky background.

\begin{figure}[!ht]
  \vspace{5mm}
  \centering
  \includegraphics[width=3.in]{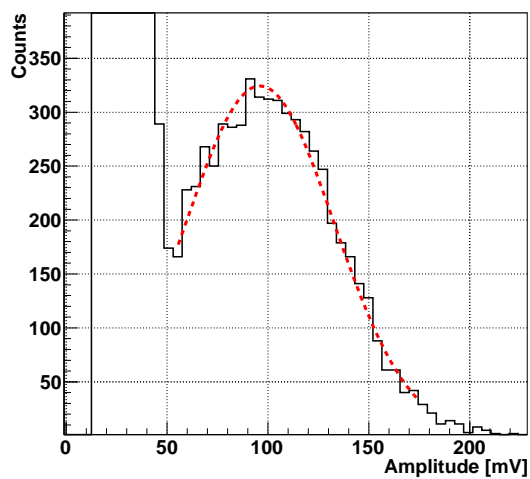}
  \caption{Typical single photoelectron pulse height distribution of the R10560. The single photoelectron peak is well separated from the pedestal and well fit with a Gaussian function (dashed red line).}
  \label{phd}
 \end{figure}

Figure \ref{phd} shows a typical single photoelectron pulse-height distribution of a R10560 that is biased at a gain of $2\cdot10^5$. The single photoelectron peak is well resolved and separated from the pedestal, which allows a solid calibration of the single photoelectron amplitude. For this measurement, the PMT was placed in a dark box and was illuminated with a pulsed light source having an average intensity of much less than one photon per pulse.

\subsection{Dynode Aging}

 \begin{figure}[!htb]
  \vspace{5mm}
  \centering
  \includegraphics[width=3.5in]{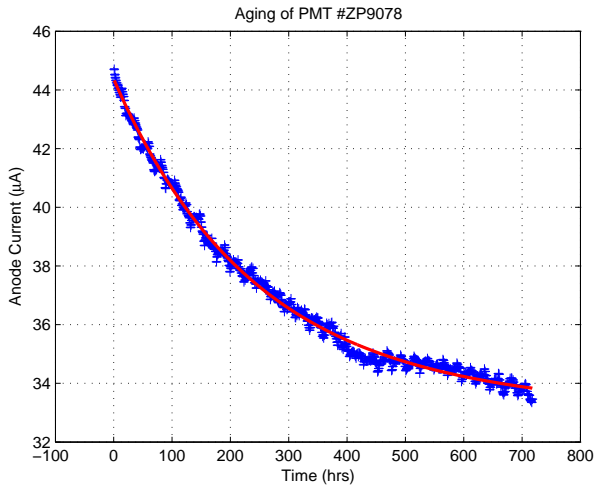}
  \caption{Gain drop of the R10560, as a result of aging. See text for explanation of the figure.}
  \label{aging}
 \end{figure}

Aging refers to the drop of gain as a function of the integral current that flows through the PMT anode. The effect cannot be neglected in IACT because the night sky background causes a steady current of a few microampere to flow through the dynode structure of the PMT. We have measured the aging during the PMT selection process on one R9800-100-20, which is the same type of PMT as the R10560, but with a borosilicate-glass entrance window instead of a UV-glass window. For the measurement the PMT was placed in a dark box and illuminated with a steady, low-level, light source resulting in an initial PMT current of 45 microampere, this is more than ten times the current during normal VERITAS observations. During the test, the PMT was biased at a fixed 1000\,V, resulting an initial gain of $2\cdot10^5$. Over a period of 700 hours, the anode current was sampled every 5 seconds. The observed drop in current as a function of time, see Figure \ref{aging}, is described by the sum of two exponential decays $I=I_0\,[ 0.25\cdot\exp(-t/\tau_{1}) + 0.75\cdot\exp(-t/\tau_{2})]$, with $I_0 = 45\,\mu$A, $\tau_1 =  247\,$ hrs, and $\tau_2 ~1300\,$days. The charge at the anode integrated over the test period is $\approx95.5$ Coulombs corresponding to roughly four years of VERITAS operation. The observed drop in gain of 25\% is well within the requirements and can be compensated by readjusting the gains on a regular basis.

\subsection{Afterpulsing}

Afterpulsing is the correlated generation of secondary pulses after the detection of single photons. Afterpulses are generated  due to, for example, residual gas atoms that are ionized by electrons in the multiplication process and that travel back to the photocathode, where they can create a large number of photoelectrons at once. If the probability becomes to high that an afterpulse is created after a photon has been detected, the accidental trigger rate of the array is dominated by afterpulses and eventually one is forced to increase the trigger threshold in order to arrive at a stable operation point and acceptable deadtimes.

 \begin{figure}[!htb]
  \vspace{5mm}
  \centering
  \includegraphics[width=2.5in]{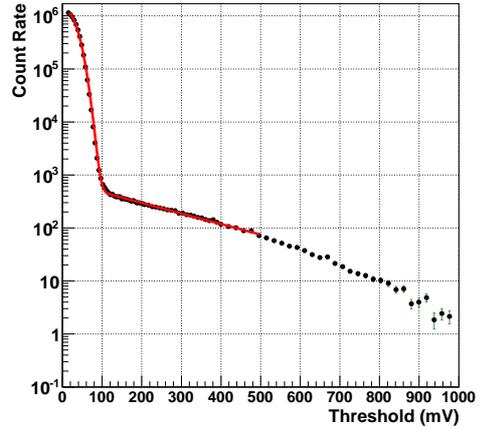}
  \caption{Count rate of the R10560 in low intensity ambient light levels as a function of the threshold of the discriminator: rate vs.\ threshold curve. The peak on the left is due to single photoelectron signals and the tail on the right is due to afterpulsing. See text for further explanations.}
  \label{afterpulsing}
 \end{figure}

We determined the afterpulsing characteristics of the investigated photon detectors with two different approaches. In the first method, the PMT is illuminated with a steady low intensity light source. The intensity of the light source is adjusted to the level where the count rate of the PMT at the single photoelectron signal level  is about one million counts per second. The afterpulsing characteristics are then determined by changing the threshold level of a discriminator and recording the count rate as a function of the discriminator level. Figure \ref{afterpulsing} shows one such rate vs.\ threshold measurement of an R10560 after background subtraction. The measured curve is fit with a function of the form $E\cdot[\exp(A+x\cdot B/C)+0.5\cdot(1+\mbox{erf}[(C-x)/(\sqrt{2}D)]]$ (red line in the figure), where the exponential term fits the afterpulsing tail and can be extrapolated down to 0\,mV, and the error function erf, fits the peak on the left side of the distribution, which is due to single photoelectron signals. From the fit the mean amplitude of the single photoelectron signals $C$ can be determined as well as the scatter of the single photoelectron amplitudes $D$.  Afterpulsing dominates the count rate above the kink in the distribution (at about 100\,mV in the figure). The data points above the kink are fit by the exponential term in the fit function. 

Averaged over all the tested R10560s, the probability is about $3\cdot10^{-4}$ that an afterpulse is generated with an amplitude that is equivalent to or larger than a five photoelectron signal. From detailed Monte Carlo simulations of VERITAS we expect that with these afterpulsing characteristics the discriminator thresholds need to be increased by less than 5\,\% in order to keep the accidental trigger rates at the array level below 10\,Hz. 

In the second method, the photomultiplier is illuminated with a pulsed, low-intensity light source and the PMT signal trace is recorded for 1.4 microseconds after each light flash. The duration of a light flash is less than 10 nanoseconds. By analyzing the recorded trace for cases in which one photon from the flash  was detected, the afterpulsing characteristics are determined by searching for subsequent pulses within the same trace, counting peak amplitudes and normalizing the counts to the number of detected photons. In addition, the peak amplitudes as a function of time after the flash allowed investigating temporal characteristics of afterpulses and identify characteristic afterpulsing \emph{times}. Both afterpulsing measurement methods yield consistent results. However, the afterpulsing probabilities are a bit lower in the second method due to the finite, 1.4 microsecond, duration of the sampling.

\subsection{Photon-Detection Efficiency}

  \begin{figure}[!t]
  \vspace{5mm}
  \centering
  \includegraphics[width=3.in]{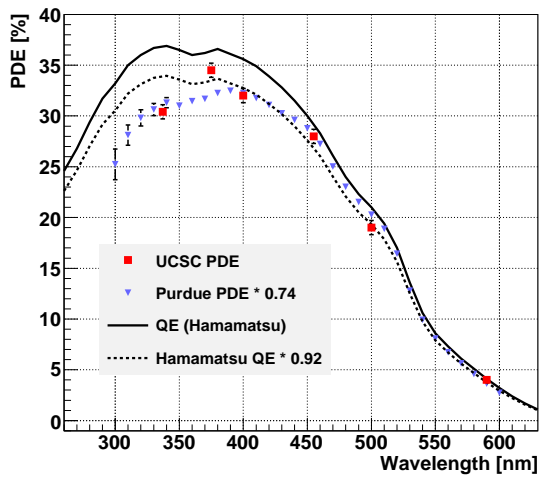}
  \caption{Photon detection efficiency, Hamamatsu QE and PDE we measured.}
  \label{pde}
 \end{figure}
 
The high quantum efficiency (QE) of the photocathode and the good photoelectron collection efficiency of $\approx90$\% are the main arguments in favor of the R10560. In particular the high collection efficiency made us chose the R10560 over the R10408, which has a collection efficiency of only 75\%.  Figure \ref{pde} shows the quantum efficiency and photon detection efficiency of one R10560 measured in different setups. The solid line shows the QE measurement by Hamamatsu, the square data points are measurements of the photon detection efficiency (PDE) performed at the University of California Santa Cruz (UCSC) with a setup described in \cite{otte}, and the downpointing triangles are relative PDE measurements performed in Purdue. The Purdue data are normalized to the UCSC measurements. The dashed line shows the Hamamatsu QE data for the same PMT scaled by 0.92 to reach approximate agreement with our own data. 

The scaling of the Hamamatsu data quantifies the photoelectron collection efficiency, which is the ratio between the PDE and the QE measurement and is 0.92 for this particular R10560. Here we did not take into account that the collection efficiency might be wavelength dependent as our measurements indicate. In addition, the differences between the data are to some extent subject to systematic uncertainties, for example, due to calibration uncertainties of the reference detectors that are used in each setup.

Folding our PDE measurement with the Cherenkov-light spectrum in the focal plane of a VERITAS telescope, we expect to detect on average 23\% of the Cherenkov photons with the R10560, which is a 35\% higher yield than the Photons XP 2970 has. 

\section{Discussion and Schedule}

Our evaluations have shown that a substantial improvement of VERITAS can be achieved if the cameras are upgraded with the R10560. The higher light yield will lower the  trigger threshold of VERITAS from about 100\,GeV down to about 70\,GeV. Furthermore, the narrower pulse shapes will allow a better discrimination of the the Cherenkov signal against NSB photons, which is advantageous during moonlight observations. 

Because the R10560 has a smaller diameter and its pin layout is incompatible with the Photonis XP2970, not only the PMT will be replaced but the preamplifier and PMT combination, which form one easy exchangeable unit in the camera. The new pixel unit consists of one Delrin tube with the same external diameter as the XP2970. The R10560 fits into the tube and is connected to a preamplifier.  The Delrin tube, PMT, and preamplifier combination is then inserted into an aluminum fitting which is inserted and fixed into the pixel supporting holey plate of the camera.  For test purposes, 14 pixels with the R10560 are operating presently in one of the VERITAS telescopes.  

The upgrade with high efficiency PMTs is funded and the PMTs are being produced by Hamamatsu over the next 10 months. They will be delivered in monthly batches of 250 PMTs. At the time of writing, May 2011, the first batch of 200 PMTs has arrived and has been tested. While each tube is tested, amongst others, for relative throughput and cathode surface homogeneity, on 10\% of the tubes detailed afterpulsing and efficiency measurements are performed. These measurements we allow us to  properly characterize the upgraded cameras and perform the necessary Monte Carlo simulations in time before the PMTs are exchanged in summer 2012.

\section{Acknowledgments}
This research is supported by grants from the US National Science Foundation, the US Department of Energy, and the Smithsonian Institution, by NSERC in Canada, by Science Foundation Ireland, and by STFC in the UK. We acknowledge the excellent work of the technical support staff at the FLWO and the collaborating institutions in the construction and operation of the instrument.

\clearpage

\end{document}